\documentclass[journal]{IEEEtran}
\usepackage{amsmath}
\usepackage{amsfonts}
\usepackage{amssymb}
\usepackage{ctable}
\usepackage{bm}
\usepackage[verbose]{cite}
\usepackage{revsymb}
\usepackage{graphicx}
\usepackage{epsfig}
\usepackage{fancybox} 

\def\x{{\mathbf x}}
\def\s{{\mathbf s}}

\def\v{{\mathbf v}}

\def\c{{\mathbf c}}

\def\q{{\mathbf q}}

\def\x{{\mathbf x}}
\def\y{{\mathbf y}}
\newcommand{\Sec}[1]{Sec.~\ref{#1}}

\newcommand{\Fig}[1]{Fig.~\ref{#1}}
\newcommand{\Figure}[1]{Figure~\ref{#1}}
\newcommand{\Eq}[1]{(\ref{#1})}
\newcommand{\MIXMAX}{mixmax}
\DeclareMathOperator{\argmax}{argmax} 

\RequirePackage[normalem]{ulem} 
\RequirePackage{color}\definecolor{RED}{rgb}{1,0,0}\definecolor{BLUE}{rgb}{0,0,1} 
\providecommand{\DIFaddbegin}{} 
\providecommand{\DIFaddend}{} 
\providecommand{\DIFdelbegin}{} 
\providecommand{\DIFaddbeginFL}{} 
\providecommand{\DIFaddendFL}{} 
\providecommand{\DIFdelbeginFL}{} 
\providecommand{\DIFdelendFL}{} 

\begin{document}

\title{Speech Separation Using Gain-Adapted  Factorial Hidden Markov Models}
\author{Martin H. Radfar$^{1,*}$, Richard.~M.~Dansereau$^{2}$ , and  Willy Wong $^{3}$\\
\mbox{$^1$ Department of Computer Science, Stony Brook University, NY, USA}\\
\mbox{$^2$ Department of Systems and Computer Engineering, Carleton University, Ottawa, Canada}
\\$^3$ Department of Electrical and Computer Engineering, University of Toronto, Canada\\
\thanks{$^*$Corresponding author: M. H.
Radfar, Email: radfar@cs.stonybrook.edu}}

\markboth{
}{M.~H.~Radfar \MakeLowercase{\textit{et al.}}: }
%

\maketitle
\begin{abstract}
 We present a new probabilistic graphical model which generalizes  factorial hidden Markov models (FHMM) for the problem of single channel speech separation (SCSS) in which we wish to separate the two speech signals $X(t)$ and $V(t)$ from a single recording of their mixture
$Y(t)=X(t)+V(t)$ using the trained models of the speakers' speech signals. Current
 techniques assume the data used in
the training and test phases of the separation model have the same loudness. In this paper, we introduce GFHMM, gain adapted FHMM, to extend SCSS to the general case in which $Y(t)=g_xX(t)+g_vV(t)$, where $g_x$ and $g_v$ are unknown gain factors. GFHMM consists of two independent-state  HMMs and a hidden node which model spectral patterns and gain difference, respectively. A novel inference method is presented using  the  Viterbi algorithm and  quadratic optimization with minimal computational overhead.  Experimental results, conducted on 180
mixtures with gain differences  from 0 to 15~dB, show that the proposed technique significantly outperforms FHMM and its memoryless counterpart, i.e., vector quantization (VQ)-based SCSS.
\end{abstract}

\vspace{2 mm}


\section{Introduction}
\DIFaddbegin \label{sec:Intro}

The human auditory system is able to
 listen and follow one speaker in presence of others. Replicating this ability by machine is one of the most challenging topics in the field of speech processing. Historically,
 Cherry \cite{Cheery1957} was the first to  introduce  this topic as ``cocktail party
problem". Later on Bregman coined the term ``computational auditory scene analysis"  (CASA)  to refer to methods that separate a desired sound from a mixture by detecting and grouping the discriminative features pertaining to the desired sound \cite{Bregman1994,Divenyi2004,Ellis06}.  Discrepancy in pitch frequencies \cite{Bregman1994,Wangbook2006,Li2006,BrownCooke1994,Virtanen2000,
CASAbrown2005} and spatial diversities \cite{Cardoso1998,Jutten1991,Common1994,Bell1995,Amari1997,model2006signal} of sounds have been widely used as  discriminative features for  separation. Limiting ourself to speech signal, pitch frequency based approaches, however, fail to exploit the vocal tract related features which play an important role in speech perception \cite{Radfar84186,SCOMradfar07}. Moreover, detecting individual pitch contours from the mixed speech is extremely  difficult \cite{HuWang2004}.  On the other hand, spatial diversity is only applicable when there are two or more sensors at the scene, a prerequisite that is not met when only a single recording of the mixture is available---the so-called single channel speech separation(SCSS).

 Complexity and uncertainty of the problem of SCSS can be well-captured using  probabilistic graphical models in which the  sources and the mixture are respectively modeled by hidden and observed random variables, connected through edges showing the conditional dependency between variables. The inference is to estimate model parameters and hidden variables that maximize the joint probability of the hidden and observed variables. Among different graphical configurations,  factorial hidden Markov models (FHMM) are well-adapted to the separation problem in the single channel paradigm \cite{Roweis2000,ReyesGomezEllisJojic2004,weiss2010speech}. A FHMM comprises of two or more independent-state HMMs, each models the probability of spectral vectors of a source. In the inference stage, FHMM decodes the hidden states of independent HMMs using a multi-layered Viterbi algorithm. For SCSS, FHMM can be simplified to factorial Gaussian mixture model (GMM) \cite{Kristjansson2004, TASLAPradfar07,TASLPReddy2007}---modeling spectral vectors using GMM instead of HMM---or factorial vector quantization (VQ) \cite{Rowies2003,Radfar065}---modeling spectral vector using VQ instead of HMM. These treatments reduce the complexity of inference in expense of losing accuracy. In addition,  there have been similar probabilistic  inference  methods for the problem of SCSS, mainly based of non-negative matrix factorization, Belief propagation, and  ICA \cite{Blouet2008,Rennie2008,WASPAA2007Schmidt,Weiss2006,schmidt06speechseparation,JangLee2003,king2010single,gaosingle,rennie2010single,starksingle,rennie2009single}.

Regardless of the applied probabilistic model, most current techniques
suffer from a fundamental shortcoming, known as gain mismatch. These techniques
assume that the data used in the modeling and  inference phases are recorded in similar
condition such that the trained model obtained from the training data set is valid
for the test speech signals. This condition, however, is not always met. For
instance, during the  test phase,  speakers may utter  test signals  louder or
weaker than when they have uttered the training data set. As such,  a mismatch may
occur between the model and test speech files. In this paper, we propose a new probabilistic graphical model, gain adapted FHMM (GFHMM) which separates the sources mixed in a general form given by $Y(t)=g_xX(t)+g_vV(t)$ where $g_x$ and $g_v$ are positive scale
factors.  GFHMM infers the hidden states and gain ratio using an iteration method consisting of Viterbi decoding and quadratic optimization. In contrast to multi-layer FHMM, GFHMM  does not add extra layers to model the gains, so it does not increase computational complexity.   We show that GFHMM improves the separation performance significantly compared to FHMM and a memoryless version  of  GFHMM  known as gain adapted vector quantization-based separation GVQ \cite{radfarmonaural}.

The rest of this paper is organized as follows. In \Sec{sec:Preliminary}, preliminary
definitions, notations, and models  used for representing signals  are given . In \Sec{sec:HMM-based scaled SCSS},  GFHMM  is described.  In \Sec{sec:recovery}, the procedure for recovering the sources from estimated GFHMM parameters is introduced.
Experimental results are reported in \Sec{sec:Experiments} where  GFHMM is compared with FHMM and GVQ.  Finally, conclusions are drawn in \Sec{sec:Conclusions}.


\section{Definitions and  Models}
\label{sec:Preliminary}


\subsection{Expressing Observation Signal Energy in Terms of the Scale Factors }

 Let $X(t)$ and $V(t),\,\, t=0,\dotsc,T-1$, be  the target and
interference speech signals. Let $g_x$ and $g_v$ be two positive real values which
represent the scale factors. The observation signal $Y(t)$ is then given by
\begin{equation}
Y(t)=g_xX(t)+g_v V(t),\qquad t=0, \dotsc,T-1 \label{Eq:latas}.
\end{equation}
where $\{g_x,g_v\}>0$  represents the associated source gains and it is assumed that
the signals have equal power before gain scaling,
$G^2=\frac{1}{T}\sum_{t=1}^TX^2(t)=\frac{1}{T}\sum_{t=1}^TV^2(t)$. From
\Eq{Eq:latas}, we obtain
\begin{equation}
g_y^2=\frac{1}{T}\sum_{t=1}^TY(t)^2=G^2(g_x^2+g_v^2)+2g_xg_v\frac{1}{T}\sum_{t=1}^T
X(t)V(t).
\end{equation}
The  minimum mean square error estimate of the observation's gain given $g_x$ and
$g_v$ is obtained by
\begin{equation}
E(g_y^2|g_x,g_v)= E(G^2)(g_x^2+g_v^2)+2g_xg_vE\Bigl(\sum_{t=1}^T X(t)V(t)\Bigr)
\end{equation}
where $E(\cdot)$ denotes the expectation operator. Since $X(t)$ and $V(t)$ are
zero-mean independent random processes, $E\Bigl(\sum_{t=1}^T X(t)V(t)\Bigr)=0$.
Hence, we obtain
\begin{equation}
E\Bigl(g_y^2|g_x,g_v\Bigr)=E(G^2)(g_x^2 +g_v^2).
\end{equation}
The probability density function of  $G^2$ is modeled by
 a Gaussian distribution with mean $G_0^2$. Thus, we obtain
\begin{equation} g_y^2= G_0^2(g_x^2+g_v^2)\label{Eq:MMSEforgains}\end{equation}
where we ignore the estimation error. Let's define the target-to-interference ratio (TIR),  which gives the energy ratio  between the target and interference,  as
\begin{equation}
\theta=10\log_{10}\frac{g_x^2}{g_v^2}. \label{Eq:SSRdefinition}
\end{equation}
From \Eq{Eq:MMSEforgains} and \Eq{Eq:SSRdefinition}:
 \begin{eqnarray*}
 g_x&=&\frac{g_y}{G_0}(1+10^{\frac{-\theta}{10}})^{-\frac{1}{2}}\\
 g_v&=&\frac{g_y}{G_0}(1+10^{\frac{\theta}{10}})^{-\frac{1}{2}}
 \end{eqnarray*}
Denoting $g(\theta)= \log_{10}\frac{g_y}{G_0}(1+10^{\frac{-\theta}{10}})^{-\frac{1}{2}}$ yeilds
   \begin{equation}
\log_{10} g_x=g(\theta)\qquad\text{and}\qquad \log_{10} g_v=g(-\theta).
 \label{Eq:gaing-SSRrelation}
\end{equation} Hence, estimating $g_x$ and $g_v$
is equivalent to estimating $\theta$ as $\frac{g_y}{G_0}$ is known in advance. Therefore, we, hereafter, focus on estimating
$\theta$.


\subsection{Log Spectral Vectors of the Observation and Sources}
 \label{sec:problemformulation}
\noindent
 In this paper, we use the log spectral vectors of the windowed speech files as the input feature. Therefore, here we present the
 notations used for representing log spectral vectors of the observation, target and interference signals.
 Let  $Y(t)$, $X(t)$, and $V(t)$ be split into $R$
overlapping frames. The log spectral vectors corresponding to the observation, target
and interference for the $r^{\text{th}}$ frame are given, respectively, by
\begin{eqnarray*}
\y^r&=&\log_{10}\Big|\mathcal{F}_D\Bigl(\{Y(t)\}_{t=(r-1)M}^{(r-1)M+N-1}\Bigr)\Big|\\
&=&[y^r(0), \dotsc ,y^r(d),\dotsc,y^r(D-1)]^{\top}\\
\x^r&=&\log_{10}\Big|\mathcal{F}_D\Bigl(\{X(t)\}_{t=(r-1)M}^{(r-1)M+N-1}\Bigr)\Big|\\
&=&[x^r(0),\dotsc,x^r(d),\dotsc,x^r(D-1)]^{\top}\\
\v^r&=&\log_{10}\Big|\mathcal{F}_D\Bigl(\{V(t)\}_{t=(r-1)M}^{(r-1)M+N-1}\Bigr)\Big|\\
&=&[v^r(0),\dotsc,v^r(d),\dotsc,v^r(D-1)]^{\top}
\end{eqnarray*}
where
  $N$ and  $M$ are frame length and frame shift, respectively, ${\top}$ denotes transpose, $\mathcal{F}_D(\cdot)$ represents the
$D$-point discrete Fourier transform, and $|\cdot|$ denotes the magnitude operator.  The relation between  $\y^r$ , $\x^r$ and $\v^r$ can be
expressed using the MIXMAX
 approximation \cite{Radfar062}. According to the MIXMAX
 approximation, the log spectrum of the observation is almost
exactly equal to the element-wise maximum of  the log spectra of the target and
interference:
\begin{equation}
y^r(d)\approx \max\bigl(\log_{10} g_x+x^r(d),\log_{10} g_v+v^r(d)\bigr)\quad d=0,
\dotsc,D-1 \label{Eq:mixmaxI}
\end{equation} or, equivalently,
\begin{equation}
y^r(d)\approx \max\bigl(g(\theta)+x^r(d), g(-\theta)+v^r(d)\bigr)\quad d=0,
\dotsc,D-1. \label{Eq:mixmax}
\end{equation}

\subsection{ Modeling of Sources Using HMMs}
\label{ssec:ProbModel}
  The parameter set of a $K$-state  HMM with the discrete state sequence
  $\q^x \triangleq(q^x_1,q^x_2,\dotsc,q^x_r,\dotsc,q^x_R)$ for  the log spectral vectors of the target
  is given by $\lambda_x(\pi^x,a^x,b^x)$ where
  \begin{displaymath} \pi^x \triangleq\{\pi^x_i\},\,\,\pi^x_i=p(q^x_1=i),\quad 1 \leq i\leq K \end{displaymath}
  \begin{displaymath} a^x \triangleq\{a^x_{ij}\},\,\,a^x_{ij}=p(q^x_{r}=j|q^x_{r-1}=i),\quad1 \leq i,j\leq K,\quad 2 \leq  r\leq R,\end{displaymath}
  \begin {displaymath}
  b^x \triangleq\{b^x_j(\x^r)\},\,\, b^x_j(\x^r)=p(\x^r|q^x_r=j),\quad1 \leq j\leq K,\quad 1 \leq  r\leq
  R,
\end{displaymath}
where $p(q^x_1=i)$ denotes the initial state probability,
$p(q^x_{r}=j|q^x_{r-1}=i)$ represents the state transition probability, and
$p(\x^r|q^x_r=j)$ represents the PDF of $\x^r$ given the HMM is in state $j$. We
assume that this PDF is modeled as a  Gaussian distribution with a  diagonal
covariance matrix given by
\begin{eqnarray*}
  p(\x^r|q^x_r=j)&=&\prod_{d=0}^{D-1}p\bigl(x^r(d)|\,j\bigr)\\
  &=&\prod_{d=0}^{D-1}\frac{\exp\bigg(\displaystyle-\frac{1}{2}\Big(\frac{x^r(d)-\mu^{j}_x(d)}{\sigma^{j}_x(d)}\Big)^2\bigg)}{\sigma^{j}_x(d)\sqrt{2\pi}}
\end{eqnarray*}
where  $\mu_x^{j}(d)$ is the  $d^{\text{th}}$ component of
the mean vector  and $\sigma_x^{2j}(d)$ is the $d^{\text{th}}$ element on the diagonal of the covariance
matrix of the $j^{\text{th}}$ state.

 Likewise, a $K$-state  HMM with
the discrete state sequence $\q^v \triangleq(q^v_1,q^v_2,\dotsc,q^v_r,\dotsc,q^v_R)$
is assigned  for  the log spectral vectors of the interference defined by
$\lambda_v(\pi^v,a^v,b^v)$ where
\begin{eqnarray*} \pi^v \triangleq\{\pi^v_\ell\},\,\,\pi^v_i=p(q^v_1=\ell),\quad 1 \leq \ell\leq K \end{eqnarray*}
 \begin{eqnarray*} a^v &\triangleq & \{a^v_{\ell k}\},\,\,a^v_{\ell k}=p(q^v_{r}=j|q^v_{r-1}=\ell),\quad1 \leq \ell,k\leq K,\quad\\&& 2 \leq  r\leq R,\end{eqnarray*}
  \begin {eqnarray*}
  b^v &\triangleq& \{b^v_k(\v^r)\},\,\, b^v_k(\v^r)=p(\v^r|q^v_r=k)\quad1 \leq k\leq K,\quad \\&&1 \leq  r\leq
  R,
\end{eqnarray*}
in which $p(q^v_1=\ell)$ denotes the initial state probability,
$p(q^v_{r}=k|q^v_{r-1}=\ell)$ represents the state transition probability, and
$p(\v^r|q^v_r=k)$ represents the PDF of $\v^r$ given the HMM is in state $k$.
Similarly, we assume that this PDF is modeled using  a  Gaussian distribution with a
diagonal covariance matrix in the form  \begin {eqnarray*}
  p( \v^r|q^v_r=k)&=&\prod_{d=0}^{D-1}p\bigl( v^r(d)|\,k\bigr)\\&=&
 \frac{\exp\bigg(\displaystyle-\frac{1}{2}\Big(\frac{v^r(d)-\mu^{k}_v(d)}{\sigma^{k}_v(d)}\Big)^2\bigg)}{\sigma^{k}_v(d)\sqrt{2\pi}}
  \label{Eq:pdfhinvector}
\end{eqnarray*}
where $\mu_v^{k}(d)$ is the $d^{\text{th}}$ component of the
mean vector and $\sigma_v^{2k}(d)$  is  the $d^{\text{th}}$ element on the diagonal of the covariance matrix
of the $k^{\text{th}}$ state.

Hence, we obtain the two HMM parameter sets $\lambda_x(\pi^x,a^x,b^x)$ and
$\lambda_v(\pi^v,a^v,b^v)$ for the target and interference, respectively. These
models are used for the  separation process  described in  \Sec{sec:HMM-based scaled SCSS}
.
\section{Gain adapted FHMM (GFHMM): model and inference}
\label{sec:HMM-based scaled SCSS}

\subsection{GFHMM }

\begin{figure*}[htb]
\begin{minipage}[b]{1.0\linewidth}
  \centering
  \centerline{ \includegraphics [width=35pc]{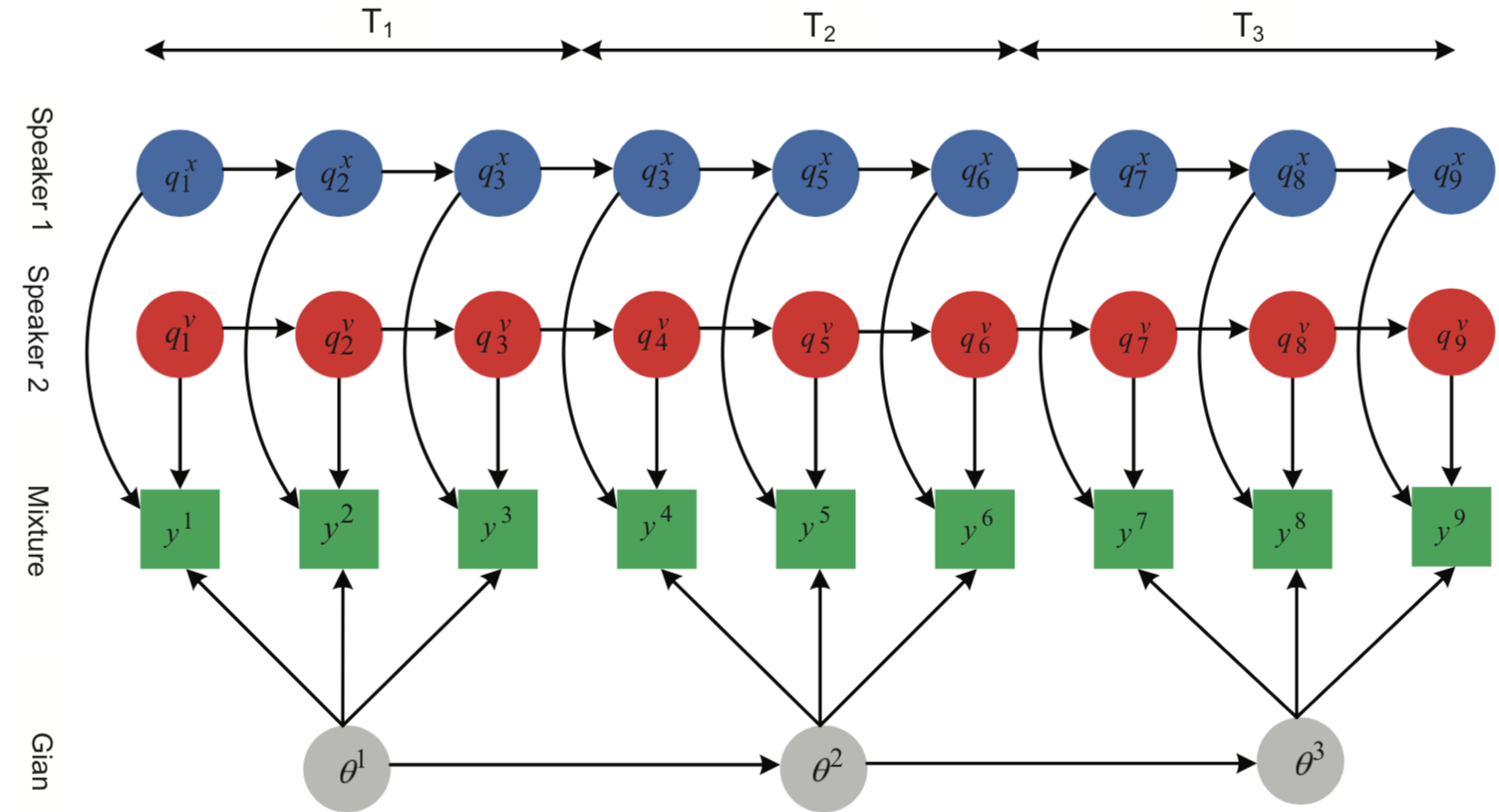}}
\end{minipage}
\caption{A graphical representation of GFHMM consisting of three hidden states modeling the log spectral vectors of two speakers ($\q^x$ and $\q^v$) and gain ratio ($\theta$) as well as an observed state modeling the mixture ($y$). Speakers' hidden states are decoded at frame level ($r=10$ msec) and gain state is decoded at mega frame rate ($T_i=2$ sec).  }
 \label{Fig:SFHMM}
\end{figure*}
We make inference using GFHMM, a graphical model  illustrated in \Fig{Fig:SFHMM}. GFHMM consists of three independent hidden layers, two HMMs correspond to speakers' spectral patterns, and one hidden node corresponds to target-to-interference ratio, $\theta$ . Our model exploits the fact that the speakers' loudness ($g_x$ and $g_v$) is almost constant within short time intervals, namely  one to two seconds. Thus, while spectral pattern decoding  is updated at each frame (10 msec), $\theta$ decoding is updated at mega frame level (1$\leq T \leq 2$sec.). Probabilistically, we wish to maximize the joint probability  of the observed signal and hidden states, given the model parameters and $\theta$. Given the observation log spectral vectors
$\y\triangleq(\y^1,\dotsc,\y^r,\dotsc,\y^R)$ and the parameter sets
$\lambda_x(\pi^x,a^x,b^x)$, $\lambda_v(\pi^v,a^v,b^v)$, and $\theta$, we aim at finding the best state sequences
$\tilde{\q}^x=(\tilde{q}^x_1,\tilde{q}^x_2,\dotsc\,\tilde{q}^x_r,\dotsc\,\tilde{q}^x_R)$ and
$\tilde{\q}^v=(\tilde{q}^v_1,\tilde{q}^v_2,\dotsc\,\tilde{q}^v_r,\dotsc\,\tilde{q}^v_R)$
 which maximize
\begin{equation}
{\tilde{\q}^x,\tilde{\q}^v}=\underset{\q^x,\q^v}{\argmax} \,\,
p(\q^x,\q^v,\y|\lambda_x,\lambda_v,\theta). \label{Eq:maximizationproblemI}
\end{equation}
For now, we assume that $\theta$ is known in advance. In the next subsection, we propose
an approach for estimating $\theta$. We solve the maximization problem in
\Eq{Eq:maximizationproblemI} using the parallel Viterbi algorithm, which is, in fact,
a two-dimensional form of the original Viterbi algorithm \cite[page 729]{Moon1999}.
To do this, we first define the variable
\begin{eqnarray*}
\delta_r(i,\ell,\theta)&=&
\underset{q^v_1,q^v_2,\dotsc,q^v_r}{\underset{q^x_1,q^x_2,\dotsc,q^x_r}{\max}}
\,\,p(q^x_1,q^x_2, \dotsc, q^x_r=i,q^v_1,q^v_2,\dotsc ,q^v_r=\ell\\&&,\y^1,\y^2,
\dotsc,\y^r|\lambda_x,\lambda_v,\theta)
\end{eqnarray*}
which is the probability corresponding to the two best  paths from the first to the
$r^{\text{th}}$ observation. For the $(r+1)^{\text{th}}$ observation, we have
\begin{equation}
\delta_{r+1}(j,k,\theta)=[\underset{i,\ell}{\max}\DIFaddbegin \ \DIFaddend
\delta_r(i,\ell,\theta)\,a^x_{ij}\,a^v_{\ell k}]b_{j,k}(\y^{r+1}|\theta)
\end{equation}
where
\begin{equation}
b_{j,k}(\y^{r+1}|\theta)=p(\y^{r+1}|q^x_{r+1}=j,q^v_{r+1}=k|\theta).
\end{equation}
Using these definitions, we now present the parallel Viterbi algorithm. It should be
noted that the parallel Viterbi algorithm, like the Viterbi algorithm, can be
implemented by applying either  probabilities directly or the log of probabilities.
We use the latter since it reduces computations (replacing multiplication by
summation) and prevents numerical instability, which should be carefully treated
since we deal with probabilities of the order of $10^{-200}$. The parallel Viterbi algorithm
in the log domain is carried out in five steps, as follows:
\begin{itemize}
\item[\textbf{1.}]Preprocessing
\begin{itemize}
\item[$\bullet$] $\hat{\pi}^x_j=\log \pi^x_j$, and $\hat{\pi}^v_{k}=\log \pi^v_{k}$,
$\quad 1 \leq j,k\leq K$
 \item[$\bullet$] $\hat{b}_{j,k}(\y^{r}|\theta)=\log b_{j,k}(\y^{r}|\theta)$,$\quad
1 \leq j,k\leq K$,\,\, $ 1 \leq r\leq R$
 \item[$\bullet$] $\hat{a}^x_{ij}=\log a^x_{ij}$, and
$\hat{a}^v_{\ell k}=\log a^v_{\ell k}\quad 1 \leq i,j,\ell,k\leq K$
\end{itemize}
 \item[\textbf{2.}]Initialization
 \begin{itemize}
\item[$\bullet$] \DIFdelbegin
$\hat{\delta}_1(j,k,\theta)=\log\delta_1(j,k,\theta)=\hat{\pi}^x_j+\hat{\pi}^v_{k}+\hat{b}_{j,k}(\y^{1}|\theta)\quad
1 \leq j,k\leq K$
 \item[$\bullet$]$\psi_r(j,k)=0\quad
1 \leq j,k\leq K$ ,\,\, $ 1 \leq r\leq R$
 \end{itemize}
 \item[\textbf{3.}]Recursion
 \begin{itemize}
 \item[$\bullet$]
 $\hat{\delta}_{r}(j,k,\theta)=\log \delta_{r}(j,k,\theta)=\underset{1\leq i,\ell \leq K}{\max}[
 \hat{\delta}_{r-1}(i,\ell,\theta)+\hat{a}^x_{ij}+\hat{a}^v_{\ell k}]+\hat{b}_{j,k}(\y^{r}|\theta)\quad
1 \leq j,k\leq K$,\,\, $2 \leq r\leq R$ \item[$\bullet$] $\psi_r(j,k)=\underset{1\leq
i,\ell\leq K}{\argmax}
[\hat{\delta}_{r-1}(i,\ell,\theta)+\hat{a}^x_{ij}+\hat{a}^v_{\ell k}]\quad 1 \leq
j,k\leq K$,\,\, $ 2 \leq r\leq R$
\end{itemize}
 \item[\textbf{4.}]Termination
 \begin{itemize}
 \item[$\bullet$]
 $P(\theta)=\underset{1\leq i,\ell \leq K}{\max}
 \hat{\delta}_{R}(i,\ell,\theta)$
 \item[$\bullet$]  ($\tilde{q}^x_R,\tilde{q}^v_R)=\underset{1\leq i,\ell\leq K}{\argmax} \hat{\delta}_{R}(i,\ell,\theta)$
  \end{itemize}
 \item[\textbf{5.}]Path backtracking
 \begin{itemize}
 \item[$\bullet$] $(\tilde{q}^x_r,\tilde{q}^v_r)=\psi_{r+1}(\tilde{q}^x_{r+1},\tilde{q}^v_{r+1})\quad r=R-1,R-2,\dotsc,1$
 \end{itemize}
\end{itemize}

In this way, we decode the two best state sequences which maximize the joint
state sequence probability and observation. The selected state sequences are then
used to build filters whereby the target and interference are estimated. This subject
will be discussed in \Sec{sec:recovery}.

\subsection{Observation Signal Probability}

In the previous subsection, we described the parallel Viterbi algorithm for decoding
the  two best state sequences. In this algorithm,
computing $b_{j,k}(\y^{r}|\theta)=p(\y^{r}|q^x_{r}=j,q^v_{r}=k,\theta)$ plays an
important role. Here, we explain how to calculate this PDF in terms of
the PDF of the target and  the PDF of the interference, and the gain factors.  In
\cite[IV.A]{TASLAPradfar07}, we obtained an approximation to
$p(y^{r}(d)|q^x_{r}=j,q^v_{r}=k,\theta)$ in terms of the PDFs of $x^r(d)$ and
$v^r(d)$ when $g_x=g_v=0$. Adding $\log_{10}g_x=g(\theta)$ and
$\log_{10}g_v=g(-\theta)$ to $\x^r$ and $\v^r$ shifts only the means of the PDFs of
$\x^r$ and $\v^r$ by $g(\theta)$ and $g(-\theta)$, respectively, and thus the PDF of
$\y$ is given by
\begin{eqnarray*}
&&p(\y^{r}|q^x_{r}=j,q^v_{r}=k,\theta)\approx
\prod_{d=0}^{D-1}\frac{1}{\sqrt{2\pi\sigma^2_{\max}(d)}}\times\\&&\exp\Bigl(-\frac{\Bigl(y^r(d)-\max
\bigl(\mu^j_x(d)+g(\theta),\mu^k_v(d)+g(-\theta)\bigr)\Bigr)^2}{2\sigma^2_{\max}(d)}\Bigr),\label{Eq:simplepdfy}
\end{eqnarray*}
where $\sigma^2_{\max}(d)$  is the variance of the source whose mean is greater than
the other---for instance, if $\mu^j_{x}(d) \ge \mu^k_{v}(d)$, then
$\sigma_{\max}(d)=\sigma^j_x(d)$. Hence,  $\hat{b}_{j,k}(\y^{r}|\theta)$ in the
log-based
 parallel Viterbi algorithm is simply obtained by
\begin{eqnarray*}
&&\hat{b}_{j,k}(\y^{r}|\theta)=\\ &-&\sum_{d=0}^{D-1}\frac{1}{2}\Bigl(\frac{y^r(d)-\max
\bigl(\mu^j_x(d)+g(\theta),\mu^k_v(d)+g(-\theta)\bigr)}{\sigma_{\max}(d)}\Bigr)^2\\&-&
\log \sigma_{\max}(d)-\frac{1}{2}\log2\pi,\label{Eq:simplepdfy}
\end{eqnarray*}

\subsection{Estimating  $\theta$ using Quadratic Optimization}
\label{sec:Experiments}

The formulas derived for the  parallel Viterbi algorithm  assume that the $\theta$ is given in advance. Here, we propose an approach to estimate the $\theta$.  We assume that  $\theta$ lies in the interval $\Theta=[\theta_{\min};\theta_{\max}]$ over which the separation of two
signals is feasible. This means that for   those $\theta$s outside the range for
 $\Theta$, the stronger source almost completely masks the weaker source, i.e. $g_x \gg g_v
\rightarrow Y(t)\approx g_x X(t)$ or vice versa. In this paper,  we set
$\theta_{\min}=-15$~dB and $\theta_{\max}=15$~dB.

\begin{figure}[htb]
\begin{minipage}[b]{1.0\linewidth}
  \centering
  \centerline{\epsfig{figure=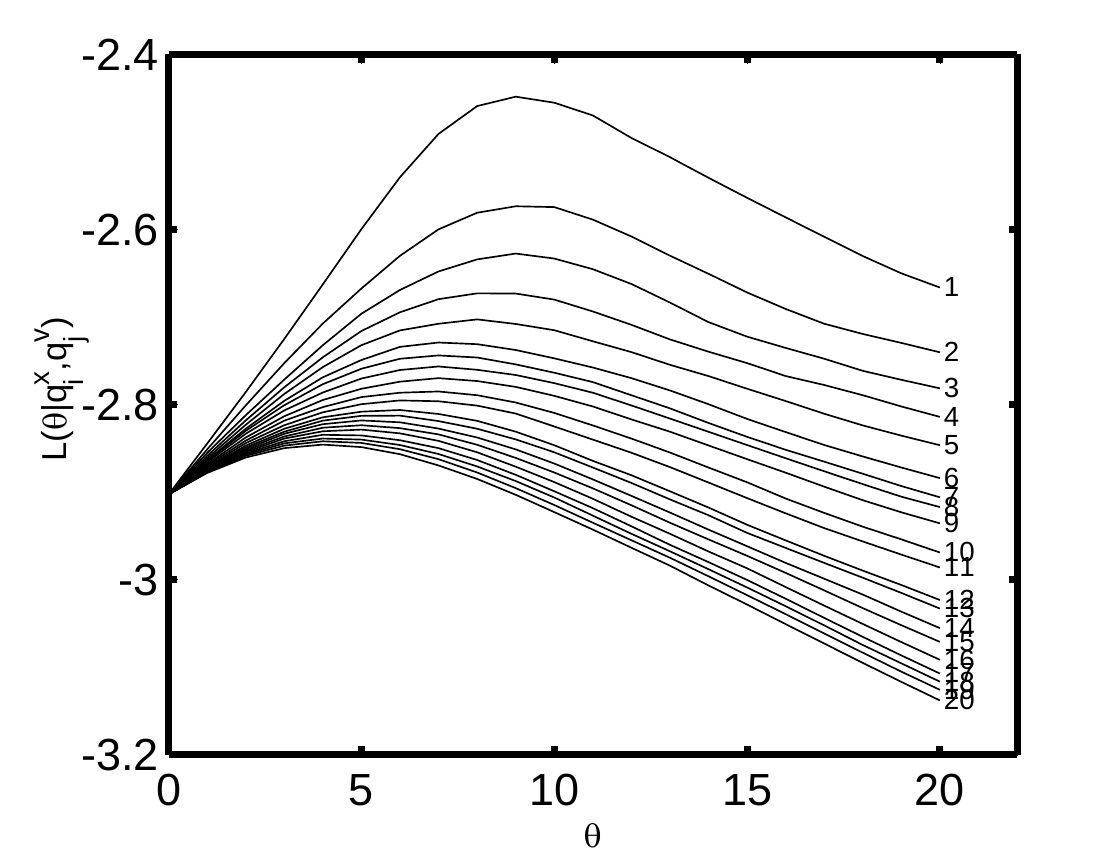,width=25pc}}
\end{minipage}
\caption{The likelihood of 20 top pairs of $(\q^x,\q^v)$ as a function of $\theta$ (dB).}
 \label{Fig:pathes}
\end{figure}

For an given arbitrary pair of $\q^x$ and $\q^v$, the joint probability $p(\q^x,\q^v,\y|\lambda_x,\lambda_v,\theta)$, \Eq{Eq:maximizationproblemI}, becomes a likelihood function of $\theta$, denoted by $\mathcal{L}(\theta|\q^x,\q^v)$. One naive approach to decode the best two pathes  and $\theta$ is to compute the likelihood function for all possible pairs in term of $\theta$ and select the pair and $\theta$  which maximize the likelihood. When implementing this procedure, we observed that $\mathcal{L}(\theta|\overset{1}{\q^x},\overset{1}{\q^v})\geq \mathcal{L}(\theta|\overset{2}{\q^x},\overset{2}{\q^v})\geq \mathcal{L}(\theta|\overset{3}{\q^x},\overset{3}{\q^v})\geq \ldots
\mathcal{L}(\theta|\overset{m}{\q^x},\overset{m}{\q^v})\geq \ldots \geq \mathcal{L}(\theta|\overset{K^2}{\q^x},\overset{K^2}{\q^v})$ for any value of $\theta$, where  $m$ in $\mathcal{L}(\theta|\overset{m}{\q^x},\overset{m}{\q^v})$ represents the rank of likelihood of a pair of states $(\q^x, \q^v)$ as decoded by the parallel Viterbi algorithm with some initial value  $\theta_0$. Moreover, we observed that $\mathcal{L}(\theta|\overset{m}{\q^x},\overset{m}{\q^v})$  approximately resembles a quadratic form. \Fig{Fig:pathes} provides an illustration of the likelihood of the 20 top pairs $(\overset{m}{\q^x},\overset{m}{\q^v})$ decoded by the Viterbi algorithm as a function of $\theta$.
As shown the likelihood associated to the best path decoded by the Viterbi algorithm is distinctly greater than the  second top for all values of $\theta$, and so forth.

\begin{figure}[htb]
\begin{minipage}[b]{1.0\linewidth}
  \centering
  \centerline{\epsfig{figure=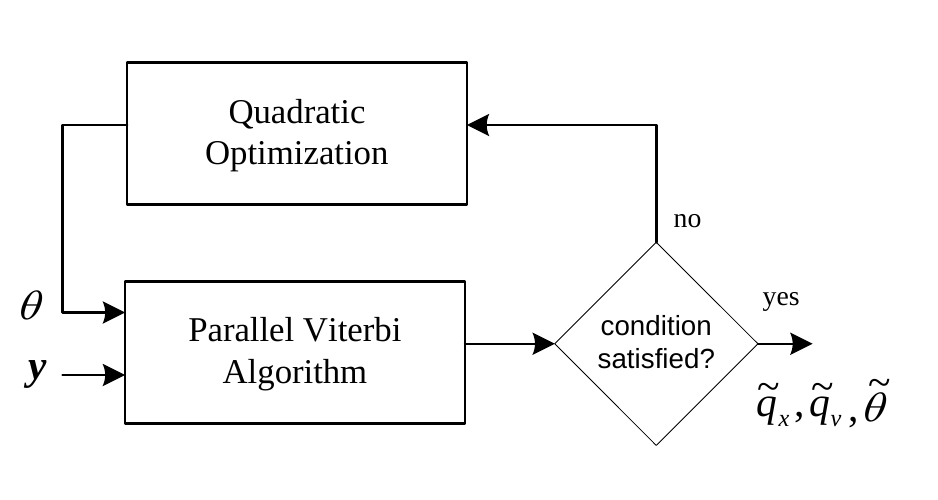,width=20pc}}
\end{minipage}
\caption{A high level block diagram of the gain adapted FHMM-based SCSS approach.}
 \label{Fig:Quadraticblockdiagram}
\end{figure}

Since $\mathcal{L}(\theta|\q^x,\q^v)$ is well-approximated by a quadratic function, $\theta$ can be readily estimated using quadratic optimization rather than performing an  exhaustive search.    \Fig{Fig:Quadraticblockdiagram} shows a block diagram of the proposed method for
decoding the best paths $\{\q_s^x, \q_s^v\}_{s=1}^S$  and $\tilde{\theta}$.  Using the decoded path  and the quadratic optimization approach, the value of $\theta$ is updated until the  maximum  of $\mathcal{L}(\theta|\q^x,\q^v)$
is reached. In the experiments, we show that the maximum is reached within two or three iterations. The procedure for estimating the maximum of a quadratic function using iterative methods can be found in \cite[page 499]{QOmethods} and \cite{bradie2006friendly}.

\DIFaddend \section{Recovering Target and  Interference  using the Decoded Parameters
} \label{sec:recovery} \DIFaddbegin

\begin{figure*}[!t]
\begin{minipage}[b]{1.0\linewidth}
  \centering
  \centerline{\epsfig{figure=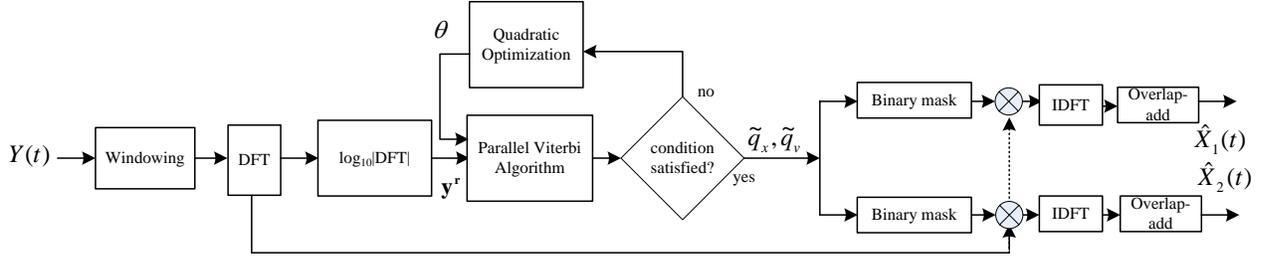,width=40pc}}
\end{minipage}
\caption{A schematic of GFHMM-based  SCSS.}
 \label{Fig:highlevel}
\end{figure*}

 In the  previous section, we proposed how to obtain the best state
sequences for trained HMMs, i.e.,  $\tilde{\q}^x$ and $\tilde{\q}^v$, and the $\theta$ which best model the mixture in a maximum likelihood sense. Here, we apply the decoded parameters to build two filters, known as binary masks, which when  applied to the mixture yields  estimates of the target and interference signals.
Using the mean vectors of the decoded states, the binary mask to
estimate the target is given by
\begin{equation} H^r_{x_{HMM}}(d)=
\begin{cases}
\,1,\,  & \mu^{\tilde{q}^x_r}_{x}(d)+\hat{g} (\tilde{\theta}) \ge \mu^{\tilde{q}^v_r}_{v}(d)+\hat{g} (-\tilde{\theta}) \\
 0 \,, & \mu^{\tilde{q}^x_r}_{x}(d)+\hat{g} (\tilde{\theta})<
 \mu^{\tilde{q}^v_r}_{v}(d)+\hat{g}(-\tilde{\theta})
\end{cases} 
\label{Eq:pbH1}
\end{equation}
$d=0,\dotsc,D-1$, whereas the binary mask for the interference is given by
$H^r_{v_{HMM}}(d)=1-H^r_{x_{HMM}}(d)$. In \Eq{Eq:pbH1}, $\mu^{\tilde{q}^x_r}_{x}(d)$
and $\mu^{\tilde{q}^v_r}_{v}(d)$ represent the $d$th components of the mean vectors
of the decoded states of the target and interference HMMs, respectively, for the
$r$th frame. The target binary mask is multiplied with the $D$-point DFT of the
 $r$th frame of the observation and then  the $D$-point inverse DFT is
applied to the resulting vector to give an  estimate of the target in the time
domain:
\begin{eqnarray*}&&\{\widehat{X}(t)\}_{t=(r-1)M}^{(r-1)M+N-1}=\\
&&\mathcal{F}^{-1}_D\biggl(H^r_{x_{HMM}}(d)\mathcal{F}_D\Bigl(\{Y(t)\}_{t=(r-1)M}^{(r-1)M+N-1}\Bigr)\biggr),
\end{eqnarray*}
$r=1, \dotsc,R$ , where  $\mathcal{F}_D(\cdot)$ and $\mathcal{F}^{-1}_D(\cdot)$ represent the $D$-point
forward  and inverse Fourier transform, respectively, and
$\{\widehat{X}(t)\}_{t=(r-1)M}^{(r-1)M+N-1}$ is the estimated target in the time
domain. Finally, the time-domain vectors are multiplied by a Hann window and
 the overlap-add method \cite{Rabiner1978} is
used to recover the  target signal. An estimated time-domain interference signal
 is obtained in a similar fashion. The entire procedure  for separating signals using GFHMM approach is shown in \Figure{Fig:highlevel}.

\DIFaddbegin

\DIFaddend \section{Experiments}
\label{sec:Experiments}
\subsection{Experimental Setup}
Speech files considered for the experiments were selected from the database presented
in \cite{Cooke2006}. The database consists of speech files of 34 speakers, each of
which uttered 500 sentences. 12 speakers were selected to form the mixtures of
female-female, male-male, and female-male pairs. Table I lists the selected speakers
and the indexes of selected speech files for evaluation. The selected speech files
were not included in the training phase. After mixing speech files,  10 female-female,
10 male-male, and 10 female-male mixtures (observations) were obtained. The speech
files were mixed at the TIRs of 0, 3, 6, 9, 12, and 15~dB  such that 180 different
observations are generated  for the experiments. One of the speech signals in each
mixture was treated as the target  while the other one as the interference. Throughout
the experiments, a Hamming window of length 32 msec with the frame shift equal to 10
msec was used to segment the speech files. Also, a Hann window was used in the
overlap-add method for synthesizing the separated speech signals. The sampling rate
was decreased to 8~kHz from the original 25~kHz in the database presented in
\cite{Cooke2006}.

For each speaker, 100 sentences were used for HMM  and VQ training (for gain adapted VQ SCSS see Appendix I). The windowed
training speech files  were transformed into the log frequency domain using a
256-point discrete Fourier transform ($D=256$), resulting in log spectral vectors of
dimension 129. For VQ modeling, the LBG VQ algorithm \cite{GershoGray1992} with
binary splitting initialization was used to construct a 64-entry codebook ($K=64$) for
each speaker. For HMM modeling, we used the  Baum-Welch method \cite{Rabiner1994} to
estimate the HMM parameters. The number of states was set to 64 ($K=64$). The initial
estimates of the HMM parameters are obtained from the VQ training. Accordingly, the
initial mean vector of each HMM state was set to a codevector and the variance of each
cluster in VQ was considered as the covariance matrix of each HMM state (we assume the
covariance matrix is diagonal). The ratio between the number of vectors in each
cluster to the total number of training vectors was used  as the initial state probability for the
corresponding state. The  Baum-Welch algorithm was terminated when   the difference
between the current and previous  log likelihoods was less than  0.00001, or the maximum number of iterations  (15) was reached.
\begin{table}[htb]
\begin{minipage} [c]{\linewidth}
  \centering
  \caption{Speech files selected for experiments from the database introduced in \cite{Cooke2006}.}
  \scalebox{0.9}{
\begin{tabular}{|c|c|c|c|c|c|c|c|c|c|c|}

  \DIFdelbeginFL 
\DIFdelendFL \DIFaddbeginFL \multicolumn{11}{l}{Type I (female-male):}
   \DIFaddendFL \\[.5ex]
 \hline spk1&1& 18& 3& 25& 6 &1& 4& 24& 3 &18\\ \hline
sen1&159& 138& 34 &130& 149 &40& 174& 162 &42& 190\\\hline spk2&4& 2& 24& 5 &7 &18 &5
&6 &11& 2\\\hline sen2&140& 182 &115 &143 &167 &133& 37& 72& 76& 76\\\hline

 \DIFdelbeginFL 
\DIFdelendFL \DIFaddbeginFL \multicolumn{11}{l}{Type II (male-male):}\DIFaddendFL \\
  [.5ex]\hline spk1&1 & 2 & 6& 17& 5& 2& 1 &3 &5& 2 \\ \hline
sen1&160& 40 &139 &174 &66& 144 &149& 35 &100& 28 \\ \hline spk2&3& 5& 17& 1 &6 &3&
6& 17& 17& 6 \\ \hline sen2&176 &196 &38 &159 &21 &76 &66 &160 &38 &22 \\ \hline

  \DIFdelbeginFL 
\DIFdelendFL \DIFaddbeginFL \multicolumn{11}{l}{Type III (female-female):}
   \DIFaddendFL \\
  [.5ex] \hline spk1&4 & 24 &7& 4& 18& 24& 25& 4 &18& 4 \\ \hline
sen1&137& 88& 63& 51& 153& 125& 124& 52& 25& 128\\ \hline spk2&18& 25& 11& 24& 25& 7&
11& 11& 7& 25\\ \hline sen2&57 &175& 72 &126&94& 129& 75& 42& 63& 40\\ \hline
\end{tabular}}
\end{minipage}
\end{table}

\subsection{Methods for Comparison}

The performance of GFHMM was compared with the gain adapted VQ (GVQ) (See Appendix I), FHMM \cite{Roweis2000}, and VQ \cite{Radfar065} based SCSS.  The comparison with the VQ-based SCSS was done to assess the performance improvement when memory (HMM-based approach) is incorporated  into the separation system.

Since  GFHMM involves an iterative stage (quadratic optimization)  and since convergence speed is an important factor for practical situations, the number of iterations required for the convergence of GFHMM was also reported.

Furthermore, to evaluate the impact of errors in the estimation of $\tilde{\theta}$ obtained by the quadratic optimization on separation performance, GFHMM was also run assuming the knowledge of the actual $\theta$ in advance (i.e., the quadratic optimization was removed and the actual $\theta$ was used).

\DIFaddend \subsection{Results}
\begin{figure}[htb]
\begin{minipage}[b]{1.0\linewidth}
  \centering
  \centerline{\epsfig{figure=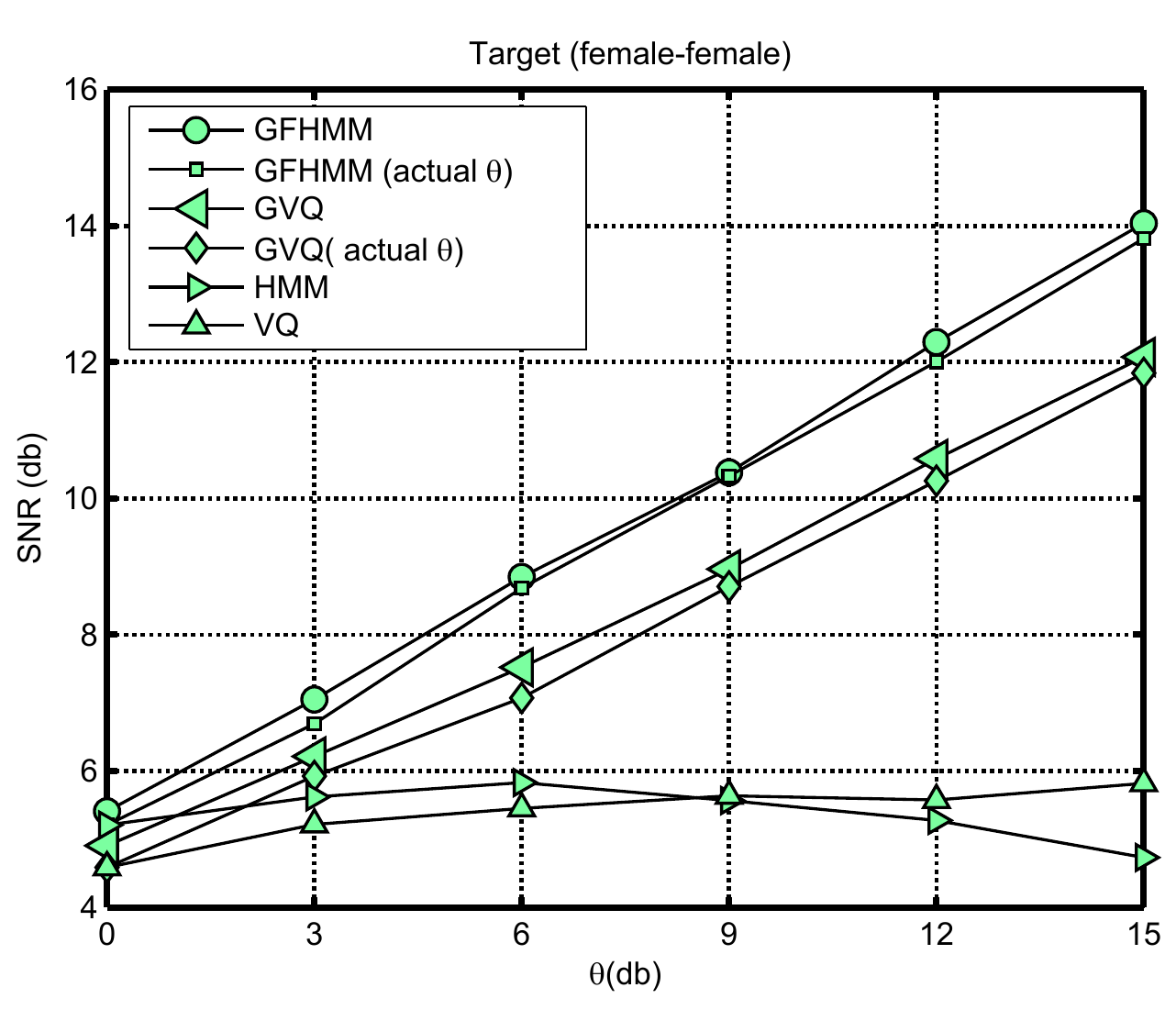,width=22pc}}
\end{minipage}
\caption{SNR versus  $\theta$    averaged over  10  separated target speech files
from female-female mixtures using GFHMM ($\circ$ line), GFHMM (with actual
$\theta$) ($\Box$ line), GVQ ($\lhd$ line), GVQ (with actual $\theta$)
($\diamondsuit$ line), HMM ($\rhd$ line), and  VQ ($\triangle$ line).}
 \label{Fig:tffsnr}
\end{figure}

\begin{figure}[htb]
\begin{minipage}[b]{1.0\linewidth}
  \centering
  \centerline{\epsfig{figure=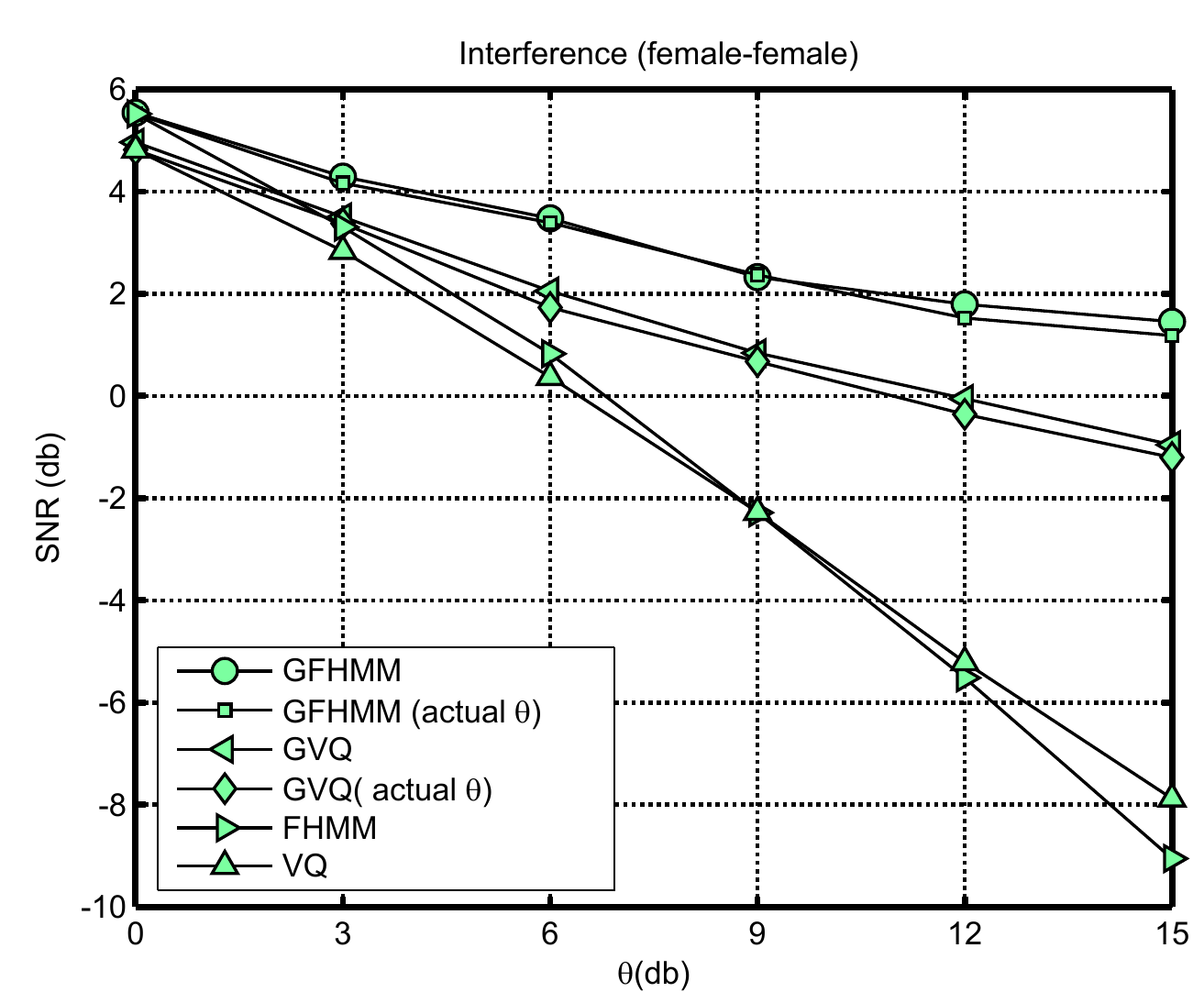,width=22pc}}
\end{minipage}
\caption{SNR versus  $\theta$    averaged over  10  separated interference speech
files from female-female mixtures using GFHMM  ($\circ$ line), GFHMM
 (with actual $\theta$) ($\Box$ line), GVQ ($\lhd$ line),
GVQ (with actual $\theta$) ($\diamondsuit$ line), HMM ($\rhd$ line), and VQ
($\triangle$ line).}
 \label{Fig:iffsnr}
\end{figure}

\begin{figure}[htb]
\begin{minipage}[b]{1.0\linewidth}
  \centering
  \centerline{\epsfig{figure=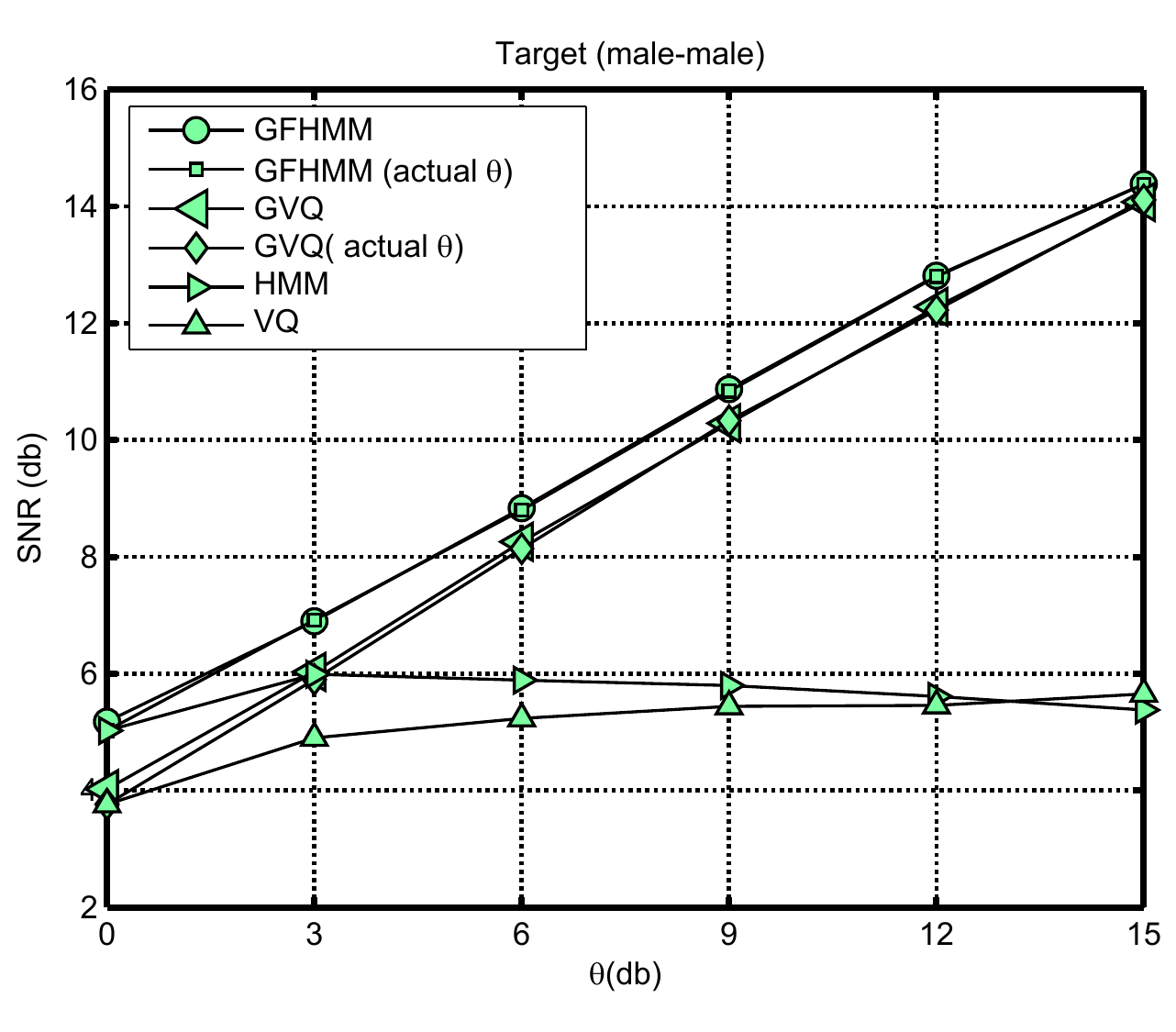,width=22pc}}
\end{minipage}
\caption{SNR versus  $\theta$    averaged over  10  separated target speech files
from male-male mixtures using GFHMM  ($\circ$ line), GFHMM (with actual
$\theta$) ($\Box$ line), GVQ ($\lhd$ line), GVQ (with actual $\theta$)
($\diamondsuit$ line), HMM ($\rhd$ line), and  VQ ($\triangle$ line).}
 \label{Fig:tmmsnr}
\end{figure}

\begin{figure}[htb]
\begin{minipage}[b]{1.0\linewidth}
  \centering
  \centerline{\epsfig{figure=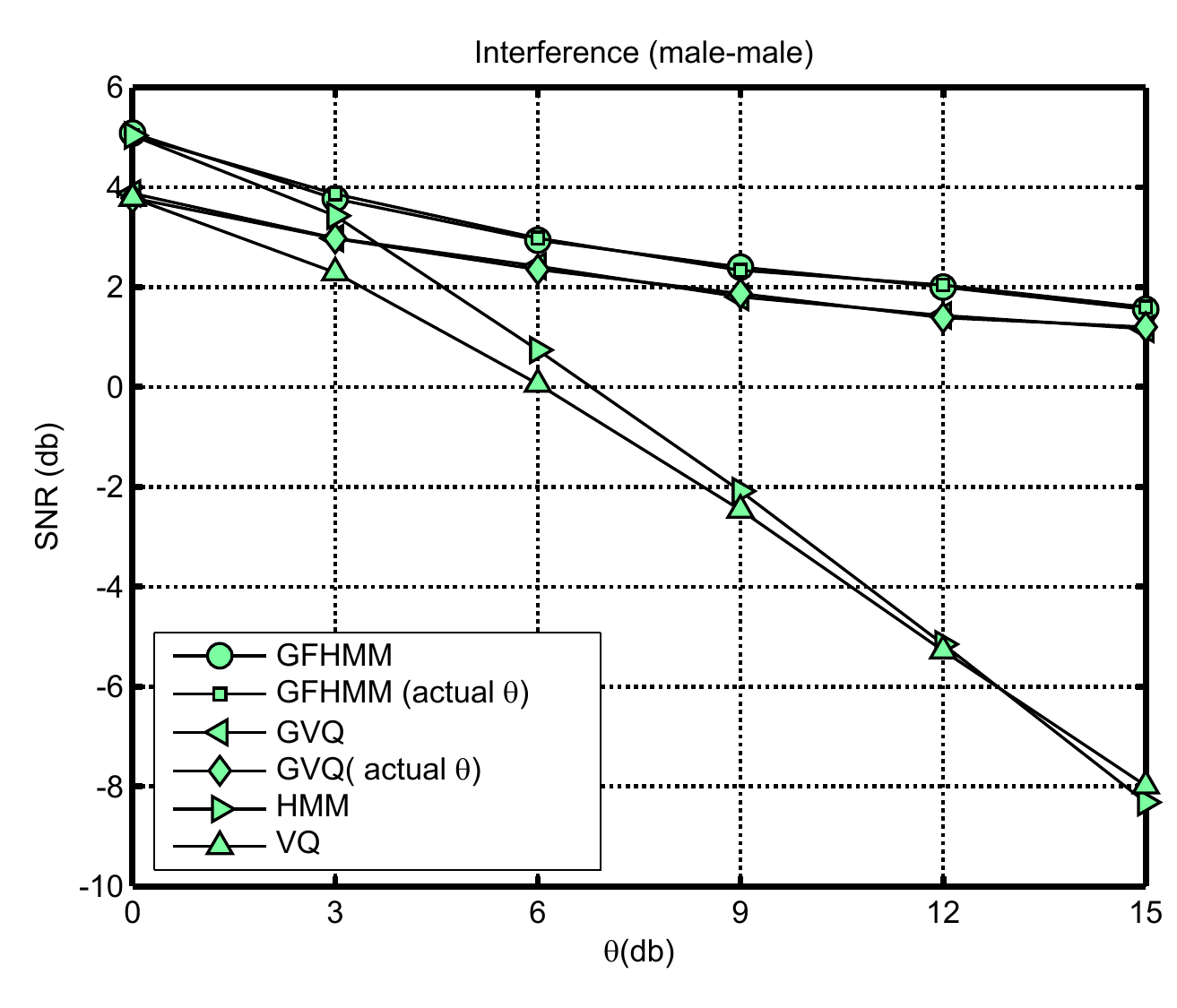,width=22pc}}
\end{minipage}
\caption{SNR versus  $\theta$    averaged over  10 separated interference speech
files from male-male mixtures using GFHMM  ($\circ$ line), GFHMM
 (with actual $\theta$) ($\Box$ line), GVQ ($\lhd$ line),
GVQ (with actual $\theta$) ($\diamondsuit$ line), HMM ($\rhd$ line), and  VQ
($\triangle$ line).}
 \label{Fig:immsnr}
\end{figure}

\begin{figure}[htb]
\begin{minipage}[b]{1.0\linewidth}
  \centering
  \centerline{\epsfig{figure=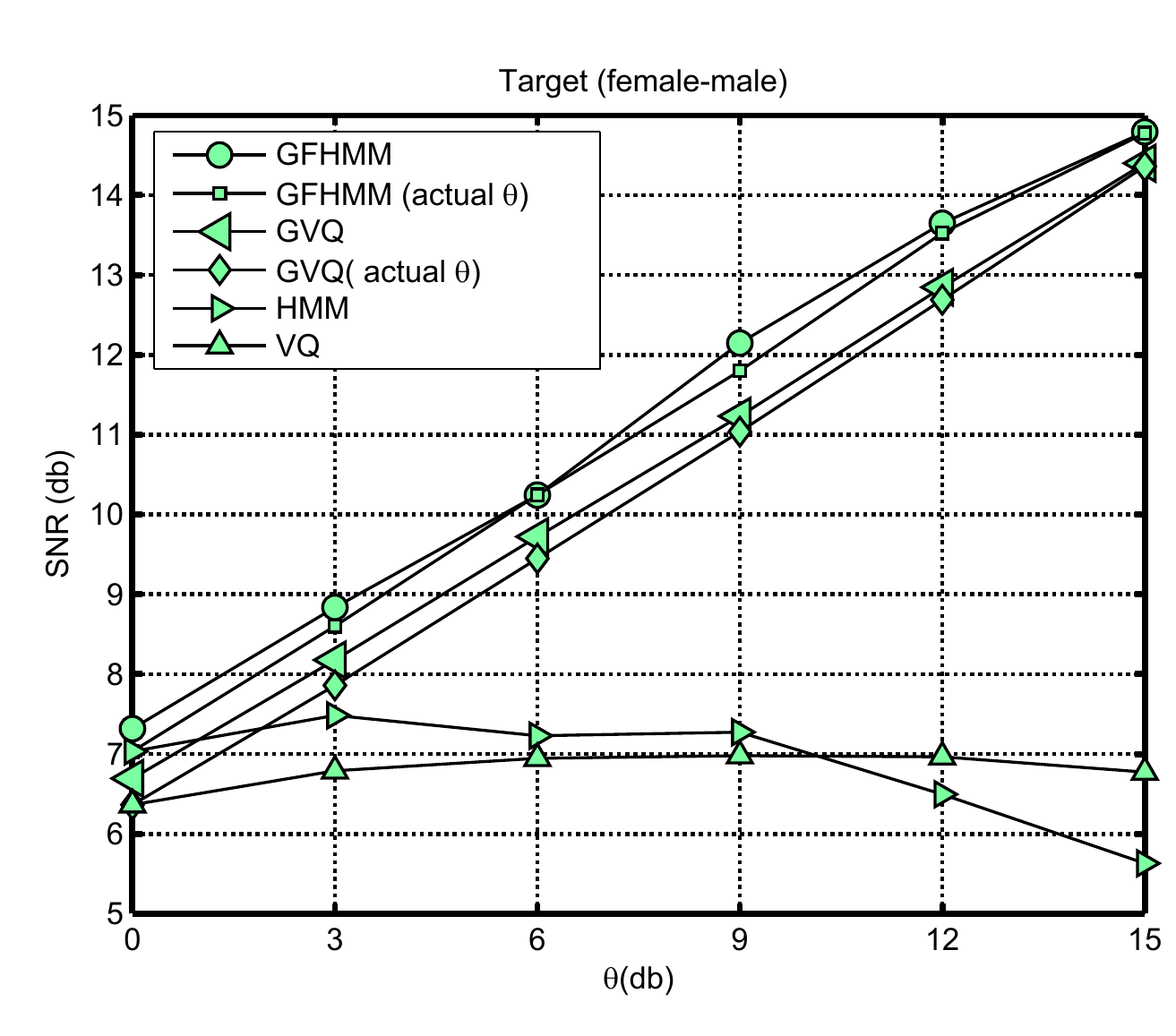,width=22pc}}
\end{minipage}
\caption{SNR versus  $\theta$    averaged over 10  separated target speech files from
female-male mixtures using GFHMM  ($\circ$ line), GFHMM (with actual
$\theta$) ($\Box$ line), GVQ ($\lhd$ line), GVQ (with actual $\theta$)
($\diamondsuit$ line), HMM ($\rhd$ line), and  VQ ($\triangle$ line).}
 \label{Fig:tfmsnr}
\end{figure}

\begin{figure}[htb]
\begin{minipage}[b]{1.0\linewidth}
  \centering
  \centerline{\epsfig{figure=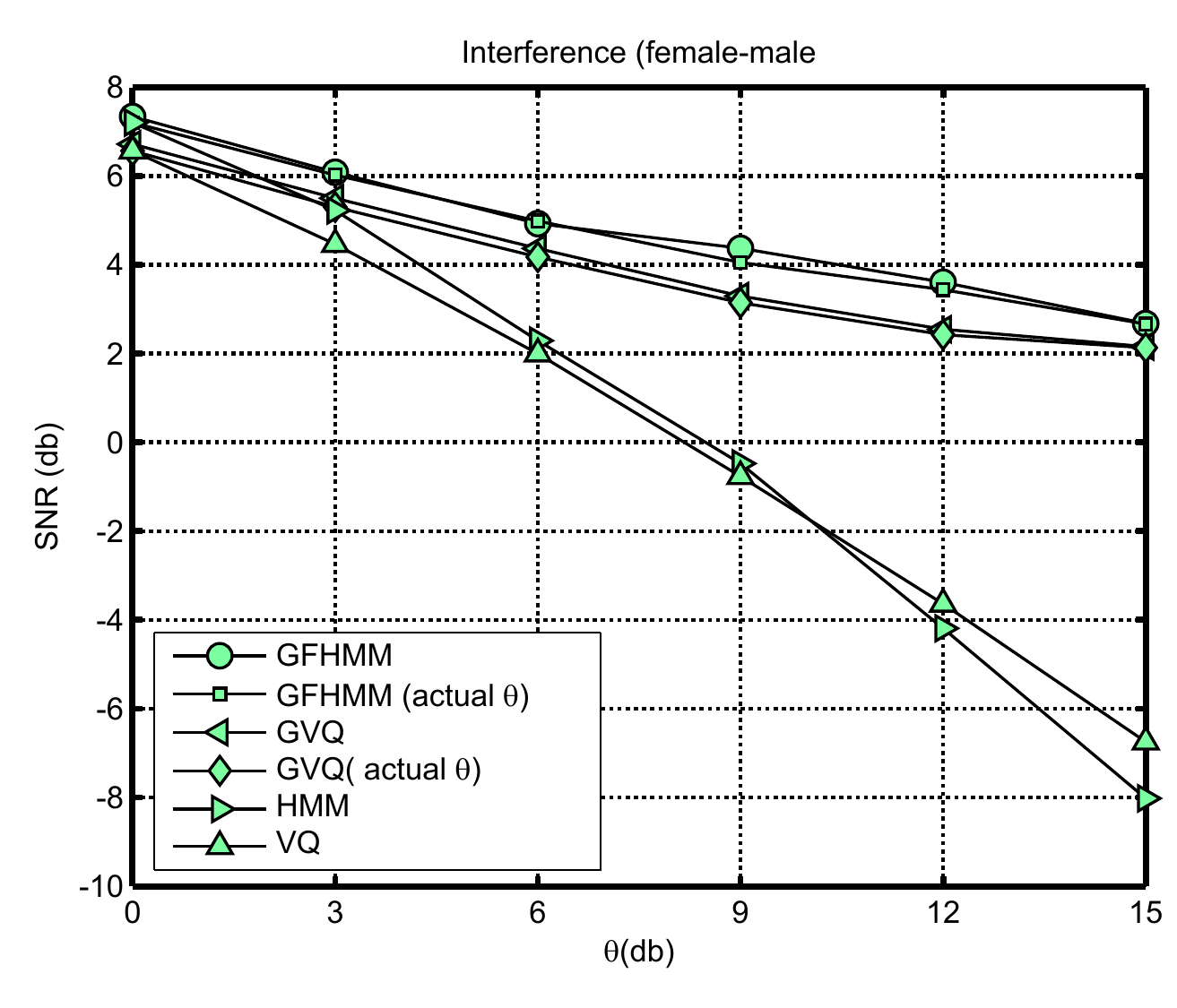,width=22pc}}
\end{minipage}
\caption{SNR versus  $\theta$    averaged over  10  separated interference speech
files from female-male mixtures using GFHMM  ($\circ$ line), GFHMM
 (with actual $\theta$) ($\Box$ line), GVQ ($\lhd$ line),
GVQ (with actual $\theta$) ($\diamondsuit$ line), HMM ($\rhd$ line), and  VQ
($\triangle$ line).}
 \label{Fig:ifmsnr}
\end{figure}

  In order to evaluate the separation performance of the proposed techniques, the signal-to-noise
ratio (SNR) between the estimated, i.e., $\hat{Z}(t)$, and original, i.e. $Z(t)$,
speech files defined by
\begin {equation}
\text{SNR}=10
\log_{10}\bigg[\frac{\sum_{t}\bigl(Z(t)\bigr)^2}{\sum_{t}\bigl(Z(t)-\hat{Z}(t)\bigr)^2}\bigg]\qquad
t = 0, 1,  \dotsc,T-1
\end{equation} was used where $Z(t)\in\{X(t), V(t)\}$.
SNR results are shown in \Fig{Fig:tffsnr}-\Fig{Fig:ifmsnr}. \Fig{Fig:tffsnr},
\Fig{Fig:tmmsnr}, and \Fig{Fig:tfmsnr} show the SNR versus $\theta$ averaged over
10  separated target speech files for female-female, male-male, and female-male
mixtures, respectively, for: GFHMM ($\circ$ line),GFHMM (with actual
$\theta$) ($\Box$ line), GVQ ($\lhd$ line), GVQ (with actual $\theta$)
($\diamondsuit$ line ), HMM ($\rhd$ line), and  VQ ($\triangle$ line). Also,
\Fig{Fig:iffsnr}, \Fig{Fig:immsnr}, and \Fig{Fig:ifmsnr} show the same respective
results, but for the separated interference signals.  From the figures, several
observations can be made which hold true for all three types of
mixtures.

The first observation is that as $\theta$ increases the SNRs for the target
signals increase as well and, on the contrary, the SNRs decrease for the interference
signals. This behavior of the SNR curves versus $\theta$ is quite expected since a
signal with higher power can be better separated from the weaker one. Comparing the
GFHMM technique with the actual and estimated $\theta$ ($\circ$ lines with $\Box$
lines), we see that the SNR results are almost the same. In fact, even slight
improvements are seen for  the scaled HMM with  $\theta$ estimated using quadratic
optimization. The same observation is also valid  for the GVQ with actual and estimated
$\theta$ ($\lhd$ lines with ($\diamondsuit$ lines). It should be noted \DIFaddbegin
that  an ideal $\theta$ for the separation process might differ from  the
 actual $\theta$ since the  actual $\theta$ is the best
choice only if the actual, rather than the model-supplied,  log spectral vectors are
used for separation.

Comparing SNR results of GFHMM and GVQ techniques ($\circ$
lines with ($\lhd$ lines), we observe that  GFHMM  outperforms the
GVQ for both target and interference signals in  all three type of mixtures.  Although the former outperforms the latter, the improvement
is not very large considering the sheer complexity of HMM when compared with VQ which
is  remarkably simpler and faster than HMM. The search complexity of  the parallel
Viterbi algorithm is
 $O(RK^3)$ \cite{Ghahramani1997} whereas  the search complexity of  the VQ technique is
 $O(RK^2)$.

We also compare SNR results obtained from  GFHMM with HMM, and GVQ with
VQ ($\circ$ and  $\lhd$ lines with $\rhd$ and  $\triangle$ lines, respectively). The results show
that the gain adapted versions of HMM and VQ  significantly outperform the
non-gain adapted ones. The improvement is quite palpable for $\theta\text{s} > 6$~ dB. The
results confirm that the model-based non-gain adapted SCSS techniques fail to separate the speech
signals when the test samples have energies substantially  different from those used
in the training data set. For all the above techniques, the separation of interference
signals at $\theta > 6$~dB is a difficult task as the separated interference
signal has  very poor quality, showing that solving this problem remains a challenge
for future studies.

\begin{figure}[htb]
\begin{minipage}[b]{1.0\linewidth}
  \centering
  \centerline{\epsfig{figure=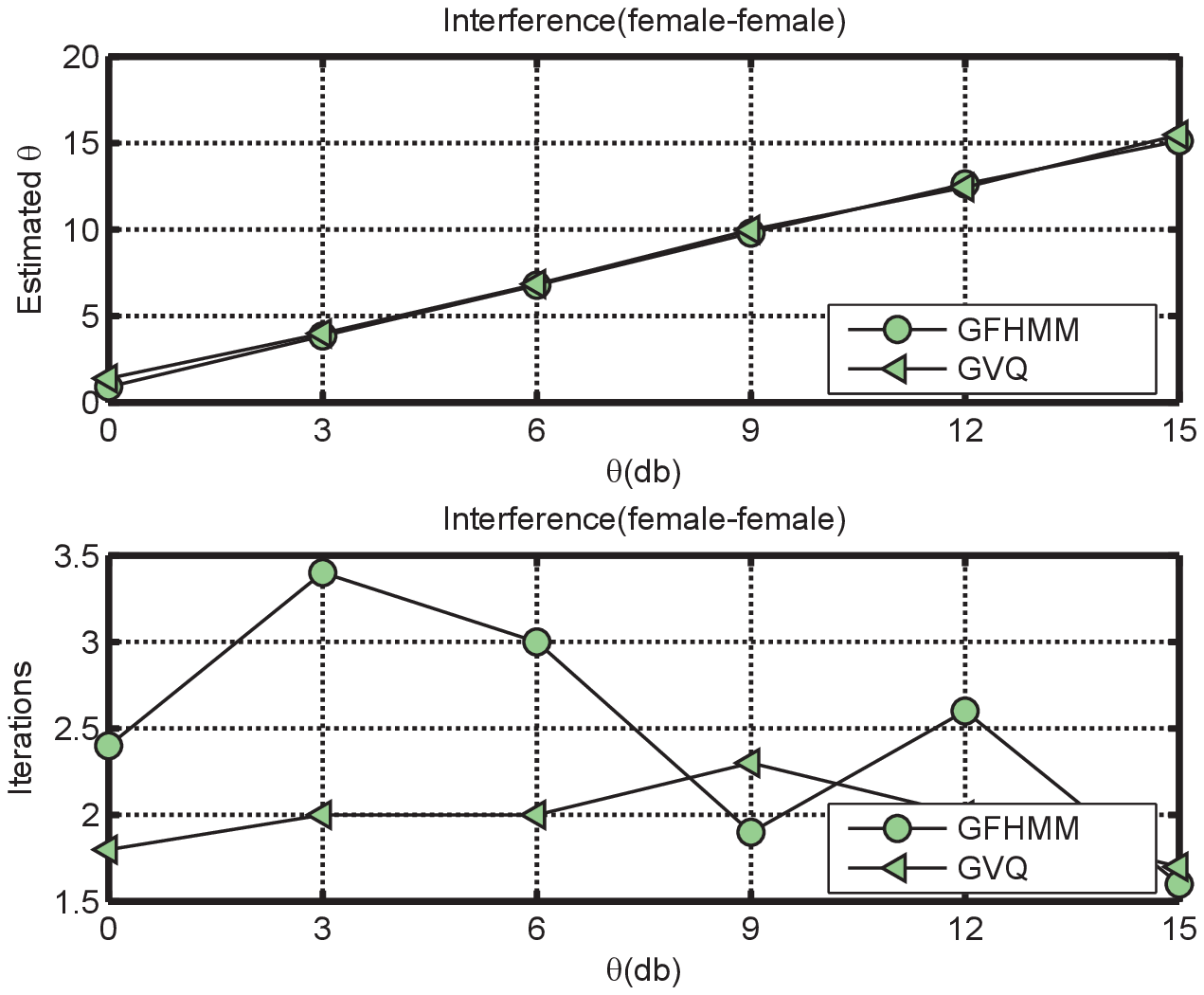,width=22pc}}
\end{minipage}
\caption{Estimated $\theta$  versus actual $\theta$  (upper panel) and number of
iterations (lower panel)  averaged over 20 separated target  and interference speech
files from female-female mixtures using GFHMM ($\circ$ line) and GVQ
($\lhd$ line).}
 \label{Fig:ffgain}
\end{figure}

\begin{figure}[htb]
\begin{minipage}[b]{1.0\linewidth}
  \centering
  \centerline{\epsfig{figure=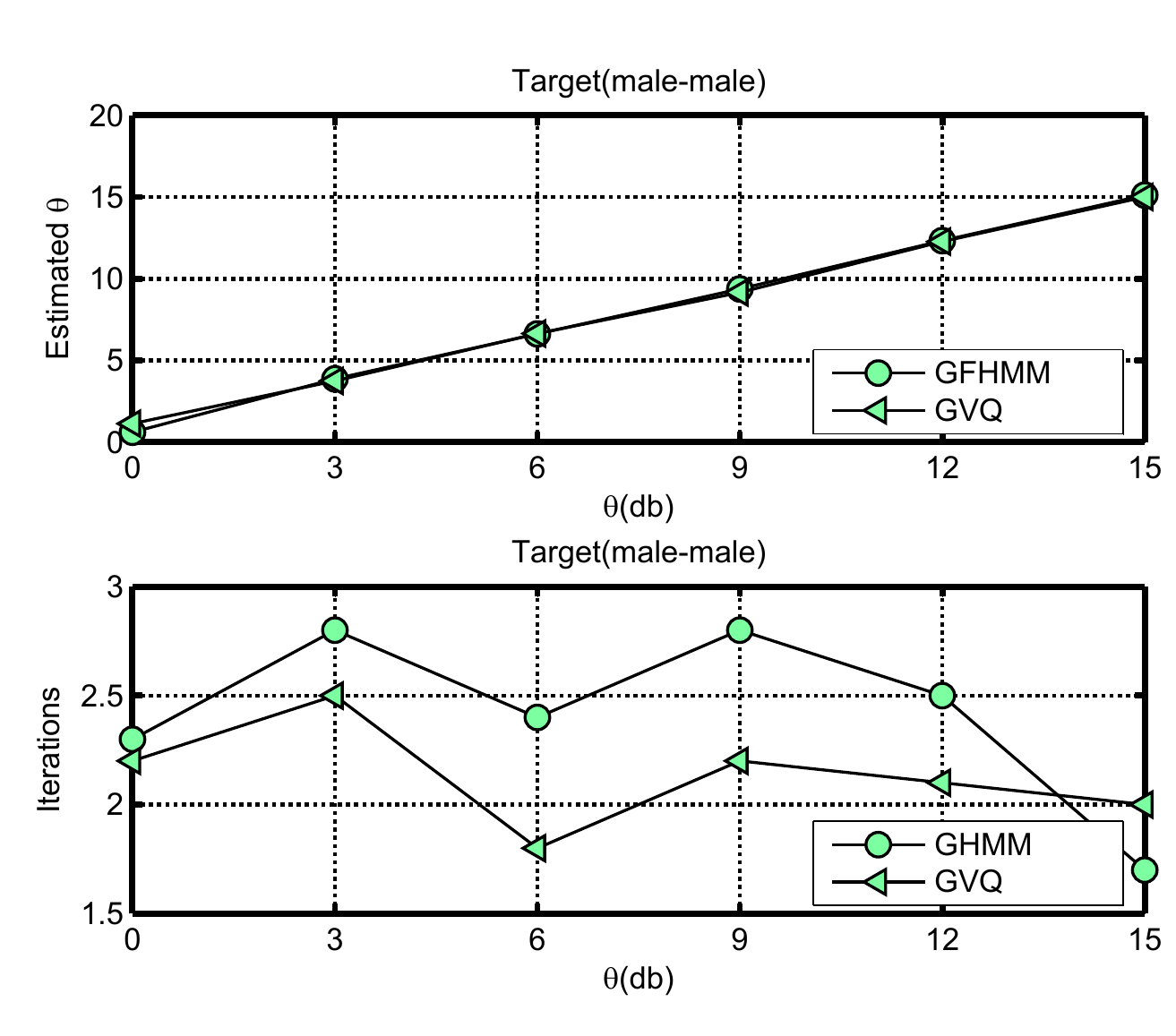,width=22pc}}
\end{minipage}
\caption{Estimated $\theta$  versus actual $\theta$  (upper panel) and number of
iterations (lower panel)  averaged over 20 separated target  and interference speech
files from male-male mixtures using GFHMM ($\circ$ line) and GVQ ($\lhd$
line).}
 \label{Fig:mmgain}
\end{figure}

\begin{figure}[htb]
\begin{minipage}[b]{1.0\linewidth}
  \centering
  \centerline{\epsfig{figure=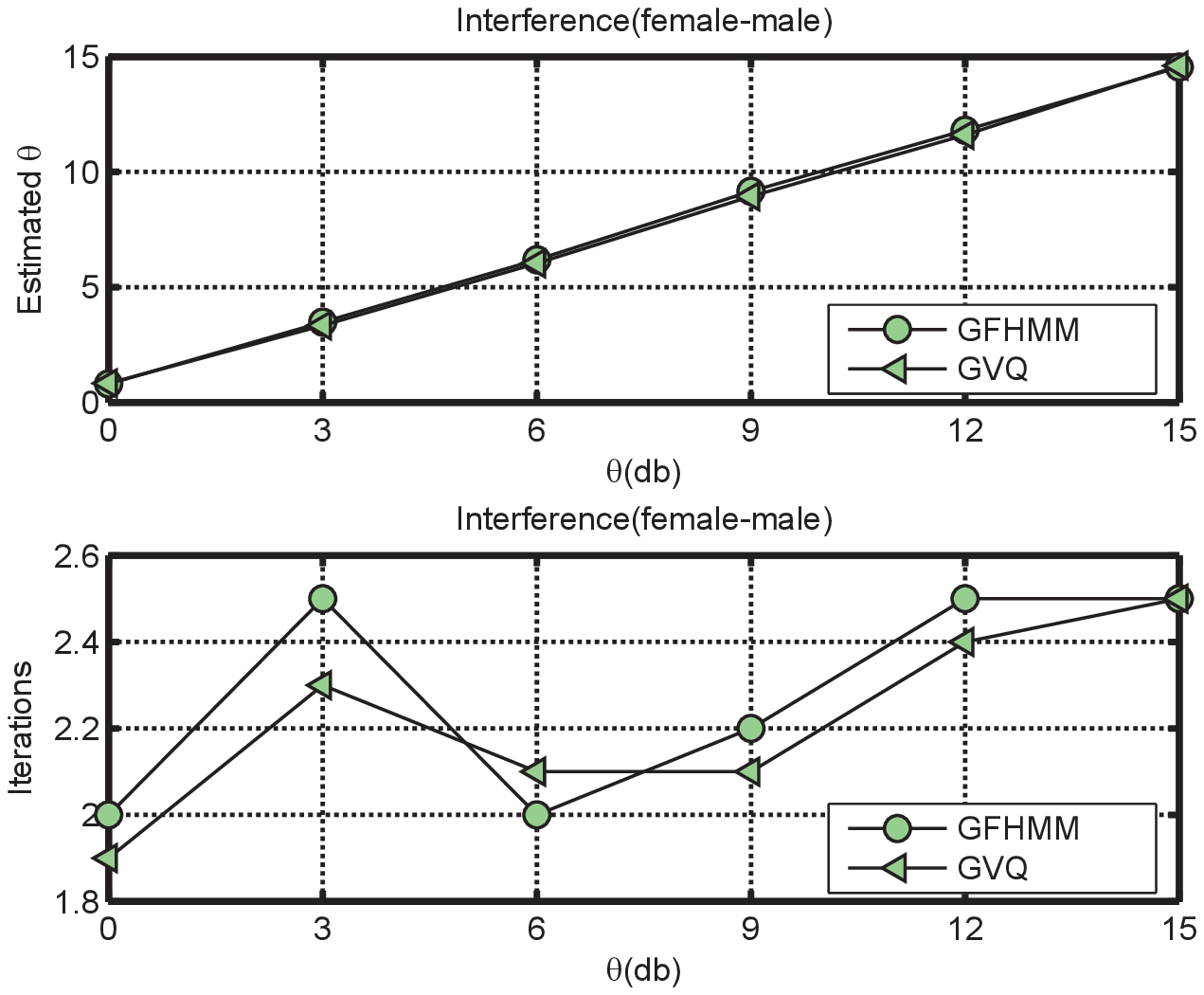,width=22pc}}
\end{minipage}
\caption{Estimated $\theta$  versus actual $\theta$  (upper panel ) and number of
iterations (lower panel)  averaged over 20 separated target  and interference speech
files from female-male mixtures using GFHMM  ($\circ$ line) and GVQ
($\lhd$ line).}
 \label{Fig:fmgain}
\end{figure}

In \Fig{Fig:ffgain}, \Fig{Fig:mmgain}, and \Fig{Fig:fmgain}, $\tilde{\theta}$ is
compared with the actual $\theta$ in the upper panels,  and in the lower panels the
number of  iterations to reach convergence is reported for GFHMM and
GVQ techniques. For both cases,  results are averaged over 10 separated speech
files. From the upper panels, one can see that estimated $\tilde{\theta}$-s are very close to the
actual $\theta$-s. Form the lower panels, it is seen that the
GFHMM converges with less than 3 iterations on average. Also, GVQ converges faster than GFHMM. The results given in
\Fig{Fig:ffgain}, \Fig{Fig:mmgain}, and \Fig{Fig:fmgain} show that the proposed
quadratic optimization approach not only well-approximate $\theta$, but also
converges very fast.


\section{Conclusions}
\label{sec:Conclusions}

 In probabilistic model based single channel speech separation, the objective is to estimate the model parameters that maximize the joint probability of the observed mixture and  the hidden variables. The exact computing of this probability is, however, intractable. Factorial hidden Markov models offer tractable approximation to the probabilistic model by decoupling states of the target and interference signals. Nonetheless, even using FHMM as a probabilistic framework, the inference becomes computationally prohibitive when more than two hidden layers are used. Accordingly, the use of FHMM for SCSS is practically limited to two-independent hidden layers (one for the target signal and one for the interference signal). Previous FHMM models assume  that speakers' loudness is the same in training and test data. In this paper we address this shortcoming by introducing a gain-adapted FHMM.  In our model, we  explicitly introduce the gain factor for the target and interference signals. In GFHMM, the number of hidden layers of the FHMM remains two, and the gain factors are estimated using quadratic optimization. This makes the computational complexity of GFHMM similar to FHMM. Our experiments show that the introduction of the explicit gain factor to the FHMM model improves the results of separation. The improvements becomes very significant when the gain mismatch between training and test signals increases. In addition to speech separation,
 GFHMM can be potentially applied to other speech processing problems such as speech enhancement and robust speech recognition where gain mismatch may reduce the performance.

\appendix[gain adapted  VQ-Based  SCSS]
 The gain adapted VQ (GVQ) \cite{radfarmonaural}  can be considered
as memoryless version  of GFHMM . In GVQ, the feature space (log spectral vectors) is partitioned into $K$  clusters  using the LBG algorithm
and the centroids of the regions are codevectors that represent  the clusters.
The goal is to find the codevectors and $\theta$ which best model the
observation. The selected codevectors and $\theta$  are then used to build
filters to recover the target and interference.  Let $\mathcal{C}_x=\{\c_x^i,\,i=1,\dotsc,K \},\,\c_x^i=\{c_x^i(d)\}_{d=0}^{D-1}$, be
a $K$-entry codebook of log spectral vectors of the target signal. Let
$\mathcal{C}_v=\{\c_v^j,\,j=1,\dotsc,K \},\,\c_v^j=\{c_v^j(d)\}_{d=0}^{D-1}$, be a
$K$-entry codebook of log spectral vectors of the interference signal.The algorithm can be explained as
follows.  Let $\s^x \triangleq(s^x_1,s^x_2,\dotsc,s^x_r,\dotsc,s^x_R)$, $s^x_r \in
\lbrace 1,\dotsc,K\rbrace$, and $\s^v
\triangleq(s^v_1,s^v_2,\dotsc,s^v_r,\dotsc,s^v_R)$, $s^v_r \in \lbrace
1,\dotsc,K\rbrace$, be the codevector index sequences for the target and interference
codebooks $\mathcal{C}_x$ and $\mathcal{C}_v$, respectively. For a given
$\theta$ and each frame, the indices of the pair of codevectors that minimize the
following cost function  are found.
\begin{eqnarray*}
&&e^r(s^x_r=i^*,s^v_r=j^*,\theta)=\underset{1\leq i,j\leq K}{\text{min}}\,\,\\
 &&\sum_{d=0}^{D-1}\Bigl(y^r(d)-\max
\bigl(c^{i}_x(d)+g_x(\theta),c^{j}_v(d)+g_v(-\theta)\bigr)\Bigr)^2.
 \label{Eq:euclidiandist}
\end{eqnarray*}
which implies that, for a given $\theta$,  the codevectors $c^{i^*}_x$
and $c^{j^*}_v$ that minimize the Euclidean distance between
 $\y^r$ and the right hand sight of \Eq{Eq:mixmax} when the original vectors $\x^r$ and $\v^r$ are replaced with the codevectors from $\mathcal{C}_x$ and $\mathcal{C}_v$.  The cost functions are summed up over all frames to get
 \begin{equation}
Q(\theta)=-
 \sum_{r=1}^{R} e^r(s^x_r=i^*,s^v_r=j^*,\theta)
\label{Eq:maximizationproblem}
\end{equation}
which is GVQ counterpart to $P(\theta)$ in GFHMM .  A similar procedure  to the one shown in \Fig{Fig:Quadraticblockdiagram} is  then carried out for the GVQ in which $P(\theta)$ is replaced with $Q(\theta)$ and the Viterbi decoding with \Eq{Eq:euclidiandist}.
Finally similar to  GFHMM, for recovering the source signals we build two binary masks using the decoded codevectors as follows
\begin{equation} H^r_{x_{VQ}}(d)=
\begin{cases}
\,1,\,  & c^{\tilde{s}^x_r}_{x}(d)+\hat{g} (\tilde{\theta}) \ge c^{\tilde{s}^v_r}_{v}(d)+\hat{g} (-\tilde{\theta}) \\
 0 \,, & c^{\tilde{s}^x_r}_{x}(d)+\hat{g} (\tilde{\theta})<
 \mu^{\tilde{s}^v_r}_{v}(d)+\hat{g}(-\tilde{\theta})
\end{cases}
\label{Eq:pbH2}
\end{equation}
  $d=0, \dotsc ,D-1$, and the binary mask for the interference signal  is  $H^r_{v_{VQ}}(d)=1-H^r_{x_{VQ}}(d)$ where $c^{\tilde{s}^x_r}_{x}(d)$ and $c^{\tilde{s}^v_r}_{v}(d)$ present $d$th components of the selected codevectors from  the target and interference codebooks, respectively,
for the $r$th frame.

\bibliographystyle{IEEEbib}
\bibliography{pitchdetection}
\vfill \pagebreak
\end{document}